# Stable DRIE-patterned SiO$_2$/Si$_3$N$_4$ electrets for electret-based vibration energy harvesters

S. Boisseau[1,2], A.B. Duret[1], G. Despesse[1], J.J. Chaillout[1], J.S. Danel[1], A. Sylvestre[2]

[1]CEA, Leti, Minatec Campus, 17 rue des Martyrs - 38054 Grenoble Cedex 9, France
[2]G2Elab, Université Grenoble 1 - Grenoble INP - CNRS, 25 Avenue des Martyrs, 38042 Grenoble, France

**Abstract**: This paper is about a new manufacturing process aimed at developing stable SiO$_2$/Si$_3$N$_4$ patterned electrets using a Deep Reactive Ion Etching (DRIE) step for an application in electret Vibration Energy Harvesters (VEH). Electrets charged by a positive corona discharge show excellent stability with high surface charge density that can reach 5mC/m² on 1.1µm-thick layers, even with fine patterning (down to 25µm) and harsh temperature conditions (up to 250°C), paving the way to new electret VEH designs and manufacturing processes.

**Keywords:** electrets, energy harvesting, charge stabilization, micro-patterning, vibration energy harvesters

Energy harvesting (EH) is a field of growing interest with a market estimated at more than $4.4 billion within ten years[1]. Many principles of EH have been investigated[2]. Among them, vibration energy harvesting has proven attractive especially when solar EH is not possible (applications in tires, building infrastructures, engines…). These systems are based on a resonant mass-spring structure which can harvest low amplitude vibrations (down to some µm)[3]. Many converters have already been developed; they use three main principles: piezoelectricity, electrostatic and electromagnetism. We focus here on electrostatic Vibration Energy Harvesters (VEH) using electrets as their polarization source. Unlike 'standard' electrostatic VEH[4], electret VEHs do not need any external polarization source; this simplifies power management electronic circuit but materials have to be further investigated during design and manufacturing[5].

Electrets are dielectrics capable of keeping quasi-permanent electric charges. They have already proven their benefits in many fields such as transducers (microphones)[6], microfluidics[6], actuators (micromotors)[6] and energy harvesting[7,8] as a high voltage polarization source that can reach more than 1000V on 100µm-thick polymer films[5] or 300V on 1µm-thick SiO$_2$-based structures[9].

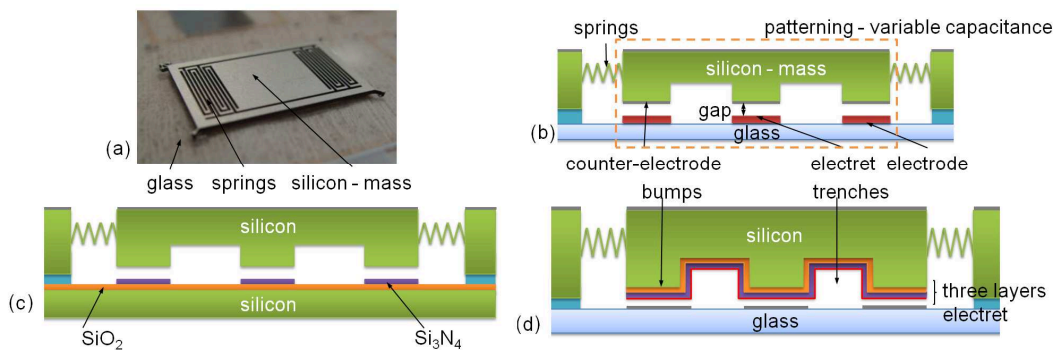

Fig 1. MEMS electret VEH a) prototype, b) diagram, c) using patterned electret from the state of the art [20] and d) using our new patterned electret.

Electrostatic energy harvesters are usually made of two or more electrodes (placed on mobile plates) which make a variable capacitance. The lower plate can be a glass wafer on which a series of electrodes is deposited. The upper plate can be a silicon wafer etched to form the mass and the springs. Some counter-electrodes are placed in front of the electrodes to make a series of capacitors. Electrets can be added on the electrodes or on the counter-electrodes. A picture and a diagram of this kind of energy harvesters are presented in Fig 1(a, b). Electret VEH output power (*P*) is directly linked to the electret surface voltage *V* and the capacitance variation *dC/dt* of the structure when it is subjected to vibrations[10]. As a first approximation, by neglecting the mechanical retroaction of the energy harvester, it appears that $P \propto V^2 (dC/dt)$. As a consequence, to harvest energy from ambient vibrations (low frequency, low amplitude), it is essential to develop a fine patterning (10µm-100µm) of electrets and electrodes that will allow a large variation of the VEH capacitance even with small displacements.

Many materials have an "electret" behavior. These materials can be separated in two categories: organic (Teflon®, Parylene, CYTOP®) and inorganic (SiO$_2$, Si$_3$N$_4$) materials. While it is quite easy to pattern organic electrets[11, 12], the same action with inorganic materials is much more difficult as it generally leads to an important charge decay through time and therefore to a limited lifetime of the VEH[13]. Nevertheless, inorganic materials such as SiO$_2$ can be really interesting for energy harvesting as they are able to keep a more important surface charge density (up to 12mC/m² [9, 14]) than organic materials which rarely exceed 2mC/m² [15, 16, 17, 18, 19, 20].

Naturally, solutions have been developed to pattern SiO$_2$-based structures and especially by IMEC (Fig 1(c))[21, 22], Sanyo and the University of Tokyo[23] and by Tohoku University[24]. Unfortunately, many of these structures cannot theoretically induce a large capacitance variation of the VEH. Indeed the capacitance is more or less constant during the relative displacement of the mobile mass compared to the fixed part as the electret surface is more or less flat: VEH output power is therefore highly limited.

To increase capacitance variation, our solution consists in etching the electret and the wafer to decrease VEH minimal capacitance. However, many experiments have shown that the main cause of charge decay is a break of the electret layer continuity. Our new electret structure is based on this statement: the electret layer is continuous while the support is patterned. This structure enables to reach a high capacitance variation which is directly linked to the trenches' deepness (Fig 1(d)).

Moreover, to get an important surface voltage with a high stability, a SiO$_2$ layer is generally not sufficient. In accordance with the state of the art[21, 25, 26], we added a Si$_3$N$_4$ (LPCVD) layer and a vapor HMDS layer on top of the SiO$_2$ layer to protect the sample from humidity. Patterned electrets are obtained from a standard 200mm p-doped silicon wafer (ρ=10 Ω.cm). The process to manufacture DRIE-patterned electrets starts with a lithography step to draw the patterning and a DRIE step to etch 100µm-deep trenches on the wafer. After cleaning, the wafer undergoes a thermal oxidization to form a continuous 1µm-thick SiO$_2$ layer. A 100nm-thick LPCVD Si$_3$N$_4$ layer is then deposited. The process continues with a thermal treatment at 450°C during 3 hours in a protective atmosphere (N$_2$). Finally, a 10nm-thick HMDS layer is deposited by vapor at 130°C. HMDS is used as a protective layer against humidity (fig 2(a)). Fig 2(a) also reports the electret geometric parameters: *e* the length of the trenches, *b* the length of the bumps, *h* the trenches' deepness and *d* the electret thickness. SEM images in Fig 2(b, c) show the patterning of the samples and the different constitutive layers. Patterning is also presented in Fig 2(d).

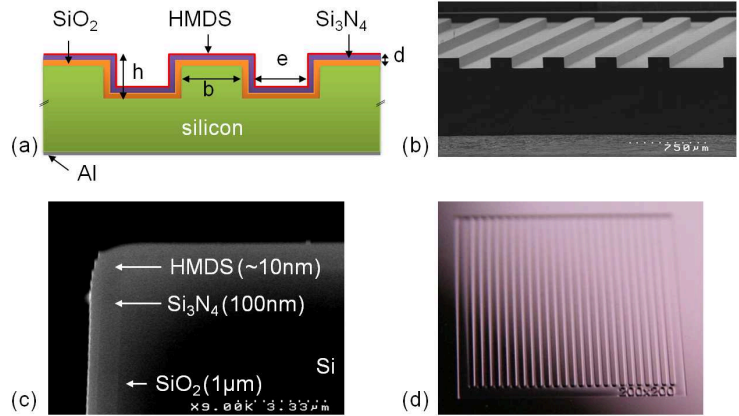

Fig 2. Patterned electrets : a) Diagram b) SEM image showing patterning c) SEM image showing the layers d) photo of one sample etched with (e, b, h)=(200µm, 200µm, 100µm)

Samples are then charged using a standard triode corona discharge (point-grid-plane) to control the surface voltage. Surface voltage $V$ is an image of charges injected into the material and can be easily measured with an electrostatic voltmeter (Trek®347). By assuming that charges are concentrated on electret surface: $V = \sigma d/\varepsilon$, with $\varepsilon$ the electret dielectric permittivity and $\sigma$ the surface charge density. This formula which is valid for a single electret layer can be modified to take the three-layer structure into account: $V = \sigma \sum_{i=1}^{3} \frac{d_i}{\varepsilon_i}$, where $d_i$ is the thickness of layer $i$ and $\varepsilon_i$ its dielectric permittivity. Electret stability is generally characterized by Surface Potential Decay (SPD), *i.e.* electret surface voltage as a function of time after charging. As VEH output power is linked to the electret surface voltage and its lifetime to the electret lifetime, SPDs are the best way to characterize electrets for an application in EH.

After charging, samples are stored in a box to protect them from air ions. Relative Humidity (RH) and temperature are also controlled (RH=20%, T=25°C) as they have generally both a large impact on the charge stability. Experiments have shown that, with our process, patterned electrets have an excellent stability even with a surface charge density up to 5mC/m² (150V of surface potential on a three-layer 1.1µm-thick electret). Some of our results are presented in Fig. 3 for different dimensions (*e*, *b*, *h*) and various initial surface voltages ($V_0$) (obtained from various grid voltages). Fig 3(a) presents the SPD of a patterned electret with an initial surface voltage of 90V (3mC/m²). This sample lost less than 1% of its initial charge in 60 days. We noted that some decay appears when exceeding ~4mC/m², as presented in Fig 3(b). This sample was charged with an initial surface voltage of 162V (5.4mC/m²). A first charge decay appears during the first 5 days when the sample loses 8% of its initial charge. The sample then stabilizes at 150V (5mC/m²) and no charge decay was observed during the next 40 days. We have also tested the effect of a post-charging thermal treatment on the electret stability. A patterned electret (Fig 3(c)) was charged with an initial surface voltage of 70V. As expected, no charge decay was observed during the first 20 days. The sample then underwent a thermal treatment of 250°C during 2H which did not affect electret stability. These results prove the stability of our patterned electrets and their validity for an application in MEMS VEH.

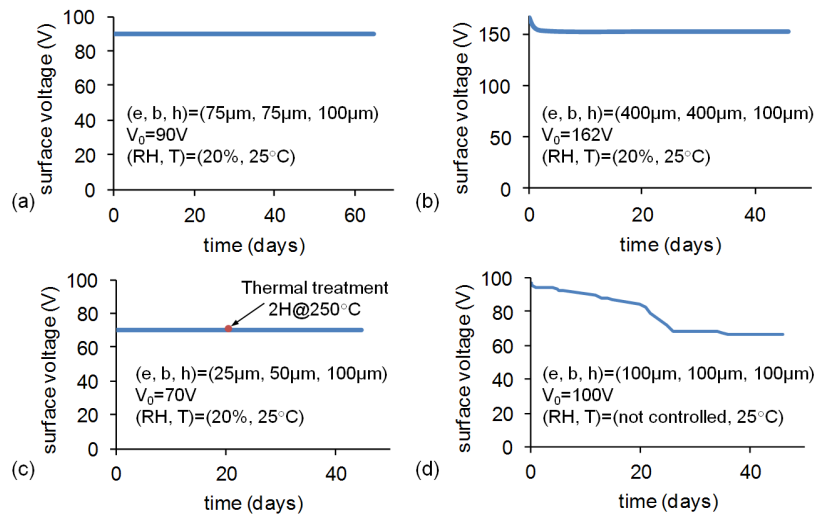

Fig 3. Example of SPDs : a) for (e, b, h)=(75µm, 75µm, 100µm) with an initial surface voltage of 90V, b) for (e, b, h)=(400µm, 400µm, 100µm) with an initial surface voltage of 162V, c) for (e, b, h)=(25µm, 50µm, 100µm) with an initial surface voltage of 70V, and d) for (e, b, h)=(100µm, 100µm, 100µm) with an initial surface voltage of 100V and with a non-controlled humidity.

Nevertheless, our experiments also showed the important impact of humidity on patterned electrets (Fig 3(d)). The sample is a patterned electret with (e, b, h)=(100µm, 100µm, 100µm) and an initial surface voltage of 100V. Unlike other samples, it was not placed in a controlled humidity atmosphere (Relative Humidity (RH) sometimes reached 70%). SPD shows that, even when using protective layers (HMDS), our patterned electrets are not able to keep their stability in non-controlled atmosphere. For VEH, the atmosphere can easily be controlled with packaging; therefore, the lack of stability due to moisture is not a problem.

To conclude, we have developed new patterned electrets allowing to induce large capacitance variations on VEH, stable through time even when working with high temperatures (up to 250°C). This new patterning solution also presents two great advantages. The first one is the charging of the sample's whole surface with the same surface voltage, limiting conduction surface phenomena between charged and non-charged zones. The second advantage is an easy manufacturing process with only 5 key steps (lithography, DRIE etching, thermal oxidization, $Si_3N_4$ LPCVD and metallization of the rear surface). We also proved the important impact of moisture on the samples even when using protective layers such as HMDS, which are well known for being good hydrophobic materials. These electrets are currently being applied to our MEMS electrostatic VEH. Studies are also being carried out to apply these new electrets to microfluidics, in order to increase surface hydrophobicity by using the double effect: surface patterning and surface potential.

**REFERENCES**

[1] IDTechEx, Energy Harvesting & Storage for Electronic Devices 2011-2021 (2011)
[2] R. J. M. Vullers et al., Solid-State Electronics 53, 684-693 (2009)
[3] C. B. Williams, and R. B. Yates, Proc. Eurosensors 1, 369-372 (1995)
[4] G. Despesse et al., Proc. sOc-EUSAI, 283-286 (2005)
[5] S. Boisseau et al., Smart Mater. Struct. 20 (2011)
[6] V. N. Kestel'man, L. S. Pinchuk, and V. A. Gol'dade, Electrets in engineering: fundamentals and applications (2000)
[7] Y. Suzuki et al., Proc. PowerMEMS (2008)

[8] H. Okamoto, T. Onuki, and H. Kuwano, Appl. Phys. Lett. 93, 122901 (2008)
[9] P. Gunther, Trans. Elec. Insul. 24 (1989)
[10] J. Boland, Y.Chao, Y. Suzuki, and Y. Tai, Proc. MEMS, 538-541 (2003)
[11] Y. Suzuki, D. Miki, M. Edamoto, and M. Honzumi, J. Micromech. Microeng. 20, 104002, (2010)
[12] D. Miki, M. Honzumi, Y. Suzuki, and N. Kasagi, Proc.MEMS, 176-179 (2010)
[13] H. Amjadi, and C. Thielemann, IEEE Trans. on Diel. and El. Ins., 3494-3498 (1996)
[14] V. Leonov, P. Fiorini, and C. Van Hoof , Trans. Diel. Elec. Insul. 13, 1049-1056 (2006)
[15] H. W. Lo, Y. C. Tai, J. Micromech. Microeng Vol. 18, 104006 (2008)
[16] Y. Sakane, Y. Suzuki, and N. Kasagi, J. Micromech. Microeng 18, 104011 (2008)
[17] R. E. Collins,  Appl. Phys. Lett. 26, 675 (1975)
[18] D. Rychkov, and R. Gerhard, Appl. Phys. Lett. 98, 122901 (2011)
[19] R. Schwödiauer et al., Appl. Phys. Lett. 76, 2612 (2000)
[20] Kashiwagi et al., Proc. PowerMEMS (2010)
[21] V. Leonov, and R. van Schaijk, Proc. electrets (2008)
[22] H. O. Jacobs, and G. M. Whitesides, Science 291, 1763-1766 (2001)
[23] Naruse et al., J. Micromech. Microeng. 19, 094002 (2009)
[24] Genda et al., JJAP, 44 (2005)
[25] Chen et al., IEEE Trans. Dielectr. Electr. Insul. 15 (2011)
[26] Amjadi et al., IEEE Trans. Dielectr. Electr. Insul. 6  (2002)